\documentclass[preprint,onecolumn,amsmath,amssymb,aps,superscriptaddress]{revtex4-1}

\usepackage{graphicx}% Include figure files
\usepackage{dcolumn}% Align table columns on decimal point
\usepackage{bm}% bold math
\usepackage{color}
\usepackage{ulem}
\usepackage{cancel}
\usepackage{multirow}
\usepackage{epstopdf}
\usepackage[english]{babel}
\definecolor{dblue}{rgb}{0, 0, 0.5}
\definecolor{dgreen}{rgb}{0, 0.6, 0}

\begin{document}

\preprint{APS/123-QED}

\title{Design rules for interfacial thermal conductance - building better bridges}

\author{Carlos A. Polanco}
\email{cap3fe@virginia.edu}
\affiliation{Department of Electrical and Computer Engineering, University of Virginia, Charlottesville,VA-22904.}

\author{Rouzbeh Rastgarkafshgarkolaei}
\thanks{Carlos A. Polanco and Rouzbeh Rastgarkafshgarkolaei contributed equally to this work.}
\affiliation{Department of Mechanical and Aerospace Engineering, University of Virginia, Charlottesville,VA-22904.}%

\author{Jingjie Zhang}
\affiliation{Department of Electrical and Computer Engineering, University of Virginia, Charlottesville,VA-22904.}%

\author{Nam Le}
\affiliation{Department of Mechanical and Aerospace Engineering, University of Virginia, Charlottesville,VA-22904.}%

\author{Pamela M. Norris}
\affiliation{Department of Mechanical and Aerospace Engineering, University of Virginia, Charlottesville,VA-22904.}%

\author{Avik W. Ghosh}
\email{ag7rq@virginia.edu}
\affiliation{Department of Electrical and Computer Engineering, University of Virginia, Charlottesville,VA-22904.}
\affiliation{Deparment of Physics, University of Virginia, Charlottesville, VA-22904}%

\date{\today}% It is always \today, today,

\begin{abstract}
We study the thermal conductance across solid-solid interfaces as the composition of an intermediate matching layer is varied. In absence of phonon-phonon interactions, an added layer can make the interfacial conductance increase or decrease depending on the interplay between (1) an increase in phonon transmission due to better bridging between the contacts, and (2) a decrease in the number of available conduction channels that must conserve their momenta transverse to the interface.  When phonon-phonon interactions are included, the added layer is seen to aid conductance when the decrease in resistances at the contact-layer boundaries compensate for the additional layer resistance. For the particular systems explored in this work, the maximum conductance happens when the layer mass is close to the geometric mean of the contact masses. The surprising result, usually associated with coherent antireflection coatings, follows from a monotonic increase in the boundary resistance with the interface mass ratio. This geometric mean condition readily extends to a compositionally graded interfacial layer with an exponentially varying mass that generates the thermal equivalent of a broadband impedance matching network.
\end{abstract}

\pacs{Valid PACS appear here}% PACS, the Physics and Astronomy
\maketitle

%\tableofcontents

\section{Introduction}

Nanostructured materials offer unprecedented opportunities for thermal management and energy conversion by enabling a wider range as well as better control of the thermal conductivity \cite{Cahill2003,Kim2007,Pop2010,Shi2015}. Interfaces are central to their performance since they are scattering centers for heat carriers whose spatial distribution can be set during fabrication and whose dispersion strength can be controlled by tailoring their physical properties \cite{Hopkins2013,Cahill2014}. Nevertheless, the full potential of this revolution is still to be seen because there is a gap between our fundamental understanding of heat flow across single and multiple interfaces and the outcome of experimental measurements \cite{Cahill2014}. For instance, while many simulations predict an enhancement of thermal conductance when a thin layer is inserted at a well bonded interface \cite{Stevens2007,Liang2011, Liang2012, English2012, Tian2012, Duda2012, Polanco2015}, only one experiment backs up that prediction so far \cite{Gorham2014}. Other experiments reporting conductance enhancement attribute the increase to a strengthening of the bonds at the boundaries \cite{Duda2013, OBrien2013, Jeong2016}. Thermal interface engineering can be critical to many technologies like integrated circuits \cite{Pop2010}, phase change memory \cite{Reifenberg2008} or high power electronics \cite{Faleev2010}. A systematic and microscopic understanding of the bridging properties of an interfacial layer would go a long way towards that goal.

\begin{figure}[tb]
	\centering
	\includegraphics[width=86mm]{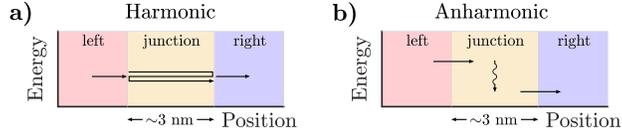}
	\caption{Interface with an added intermediate layer or junction (bridged interface) {\bf a)} In the harmonic limit, the layer behaves like an impedance matcher that increases transmission by constructive interference while reducing the number of conducting modes due to energy-momentum conserving constraints. {\bf b)} In the anharmonic limit, the layer behaves as a bridge for phonon down and up conversion that increases the chances of phonons crossing the interface.}
	\label{fig_intro}
\end{figure}

Adding an intermediate layer to a well-bonded interface can enhance the conductance in two different ways. In the harmonic limit, the layer could act as an impedance matching waveguide (Fig.~\ref{fig_intro}a) that reduces phonon reflection by destructive interference, similar to an antireflection coating \cite{Polanco2014}. Such a complete quenching of reflection occurs at a single frequency where the layer thickness can function as a quarter wave plate. In the anharmonic regime on the other hand, the layer can act as a bridge that facilitates frequency up and down conversion and increases the chances of phonons crossing the interface \cite{English2012} (Fig.~\ref{fig_intro}b). 

The contribution of each individual effect to the total enhancement has not been systematically explored on the same material system. Neither is there a clear criterion to choose the properties of the layer to maximize the conductance. In a 1D harmonic crystal for instance, we have established that a conductance maximum occurs when the impedance of the layer is the geometric mean of the contact impedances \cite{Polanco2014}, even in presence of incoherent interface scattering. This thumb-rule persists for all lengths except the extreme limit of a single atom, where the mean generates a resonance that lies beyond the cut-off frequency and the system is forced to choose an arithmetic mean instead. However, this result has not been extended to multiple dimensions and crystal structures. A similar gap exists when phonon-phonon interactions are included, where it was proposed that the maximum conductance happens when the layer's density of states (DOS) maximizes its overlap with the contact DOSs. This argument leads to two different criteria to obtain the maximum: 1) choose the atomic mass of the layer close to the arithmetic mean of the contact masses \cite{English2012} and  2) choose the Debye temperature of the layer as the geometric mean of the contact Debye temperatures \cite{Liang2012}. This unresolved discrepancy once again reveals our lack of understanding of the role played by the inserted layer for a real multidimensional physical system with complex modes, symmetries, and scattering events.

In this paper we compare the enhancement of conductance in the harmonic and anharmonic limits and demonstrate the dominant role of anharmonicity (Sec.~\ref{secharvsanh}). We show that adding an intermediate layer can go either way by increasing or decreasing the conductance when phonon transport is restricted to the harmonic regime (Sec.~\ref{secharmonic}). In this limit, the conservation of energy and momentum constrain the number of available transport channels, so the increase in average transmission per channel must compete with the loss in the number of transport channels. When anharmonicity is added (Sec.~\ref{secanharmonic}), phonon-phonon interaction relaxes the conservation constraints and decouples the boundaries. Maximizing the conductance becomes equivalent to minimizing the sum of individual boundary resistances. For our particular system, where only mass changes are considered, we show that the maximum happens when the layer mass is close to the geometric mean of the contact masses. As explained earlier, this result would be expected for 1-D coherent phonon transport at a single frequency. The surprise however is that the geometric mean ends up winning even for a 3-D crystal with broad-band phonon transport across modes and polarizations, including anharmonic and diffusive interactions. We can hypothesize that a bridging layer can in fact be a matching layer if we compositionally grade it so each slice has an acoustic impedance that is the geometric mean of its immediate nearest neighbors. The tendency of the geometric mean to favor the lower impedance of the pair mathematically translates to an exponentially
varying spatially dependent impedance, with an exponent set by the logarithm ratio of the two impedances at either end of the layer.

\section{Harmonic vs. Anharmonic Enhancement of $G$}
\label{secharvsanh}

\begin{figure}[htb]
	\centering
	\includegraphics[width=86mm]{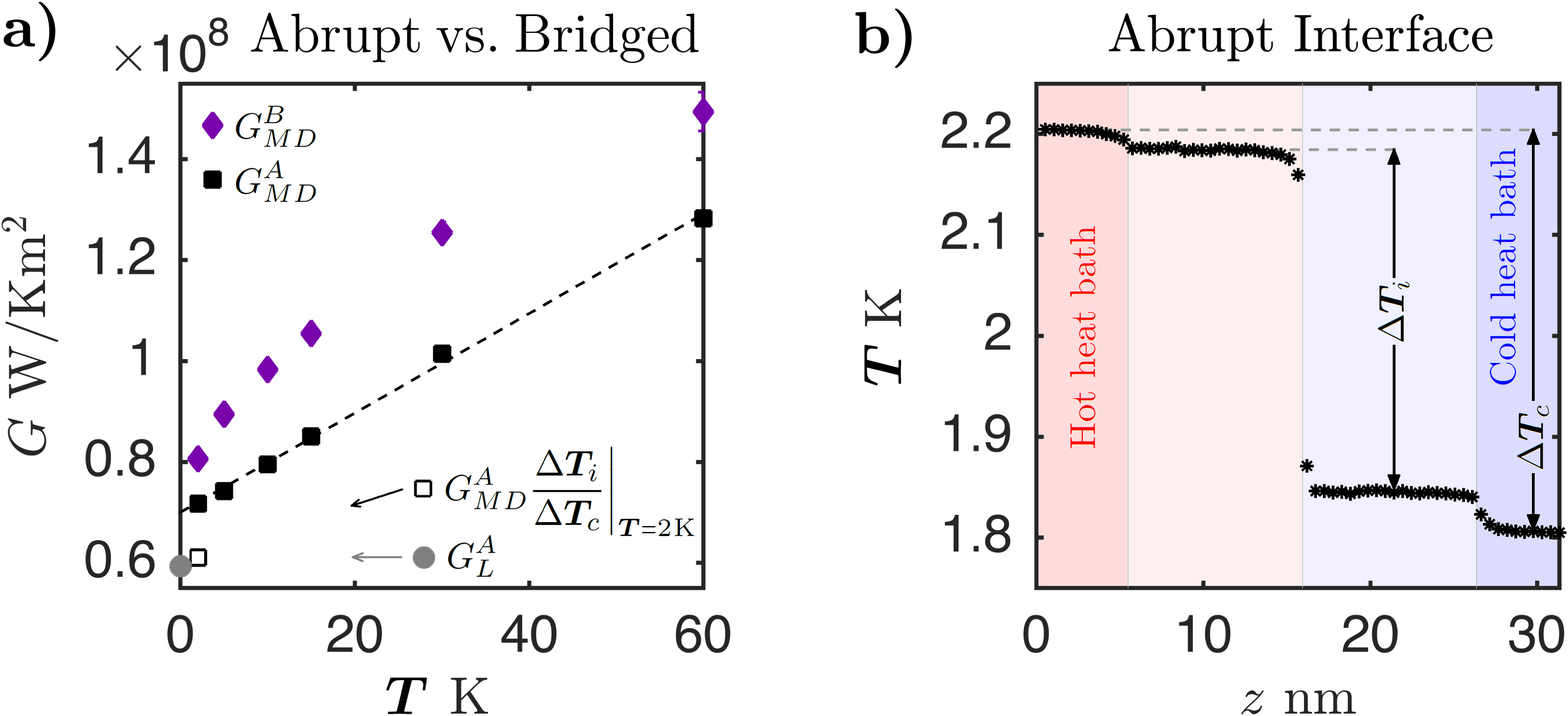}
	\caption{{\bf a)} Conductance of the abrupt interface $G^A_{MD}$ and maximum conductance of the bridged interface $G^B_{MD}$ vs. temperature for a FCC crystal calculated from NEMD. The rapid increase of $G^B_{MD}$ at low $\boldsymbol{T}$ highlights the important role of anharmonicity enhancing $G^B_{MD}$ relative to $G^A_{MD}$. {\bf b)} Temperature profile of the abrupt interface from NEMD. This profile allows us to include the contact resistances into the NEMD conductance (Eq.~\ref{equGLGMD}), which shows excellent agreement with our Landauer calculations $G^A_L$.} 
	\label{fig_DeltaGmaxvsT_LJ}
\end{figure}

Interface thermal conductance or thermal boundary conductance is defined as the ratio between the heat flux crossing an interface over the temperature drop across it, $G=q/\Delta \boldsymbol{T}$. Within the Landauer formalism, the conductance between two contacts at thermal equilibrium can be expressed as \cite{Jeong2012} 
\begin{equation}
G=\frac{1}{A}\int_0^\infty \frac{d\omega}{2\pi}\hbar\omega\frac{\partial N}{\partial \boldsymbol{T}} MT \xrightarrow{\hbar\rightarrow 0} \frac{k_B}{2\pi A}\int_0^\infty d\omega MT,
\label{equGQ}
\end{equation}
where $A$ is the cross sectional area, $\hbar\omega$ is the phonon energy, $N$ is the Bose-Einstein distribution, $k_B$ is the Boltzmann constant, $M$ is the number of available propagating channels, which we call modes, and $T$ is the average transmission per mode. In a bulk material, each mode is a 1D subband generated by a particular polarization and a transverse wavevector, which gives rise to a quantum of conductance \cite{Datta2005}. The factor $MT$ represents the sum of all the possible transmissions between the modes on the left and right contacts. This factor can be calculated using Non-Equilibrium Green's Functions (NEGF) as $MT=\text{Trace}\{\Gamma_l G_r \Gamma_r G_r^\dagger\}$, with $G_r$ the retarded Green's function describing the propagation of phonon waves in the channel, and $\Gamma_{l,r}$ the broadening matrix for the left (l) and right (r) contacts \cite{Mingo2003,Datta2005,Wang2008}. To compare the conductance from Landauer formalism with that from Non-Equilibrium Molecular Dynamics (NEMD), we need to take the classical limit of the Bose-Einstein distribution (Eq.~\ref{equGQ} with $\hbar\rightarrow 0$) and we need to subtract the contact resistance (Appendix~\ref{App_CR} and Fig.~\ref{fig_DeltaGmaxvsT_LJ}b). This value should be the limit of the NEMD conductance as temperature tends to zero. 

\begin{figure}[ht]
	\centering
	\includegraphics[width=86mm]{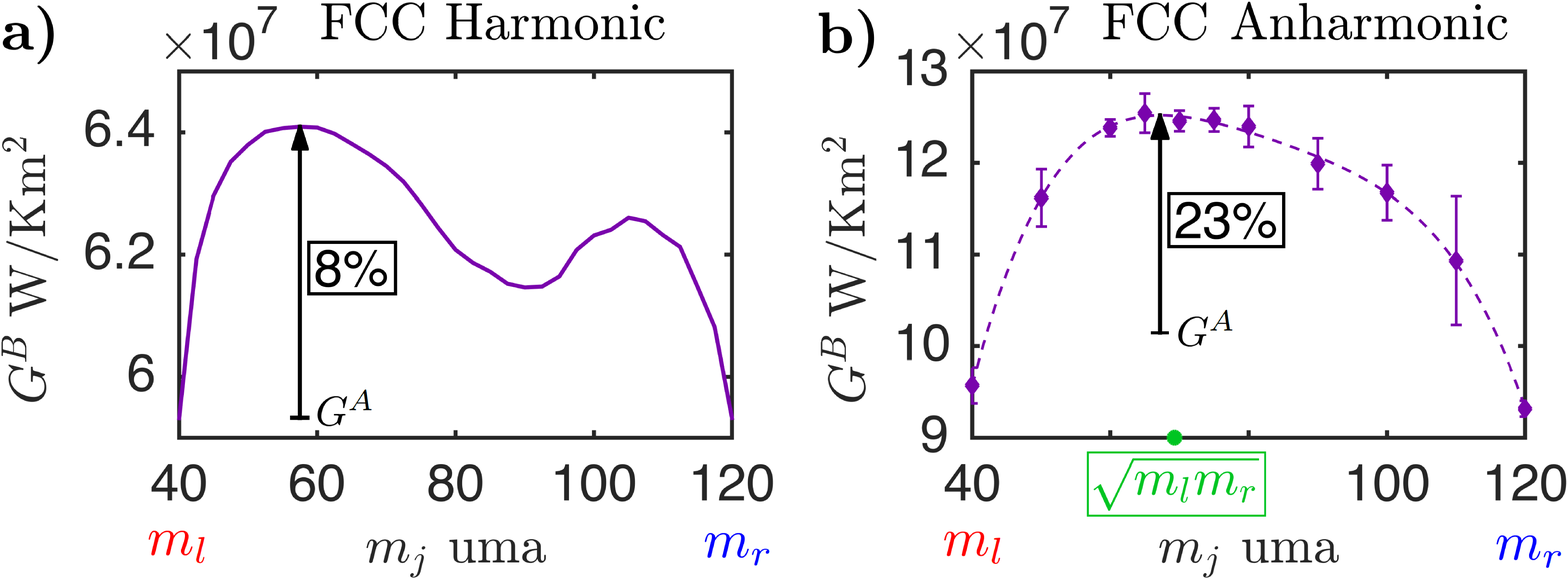}
	\includegraphics[width=86mm]{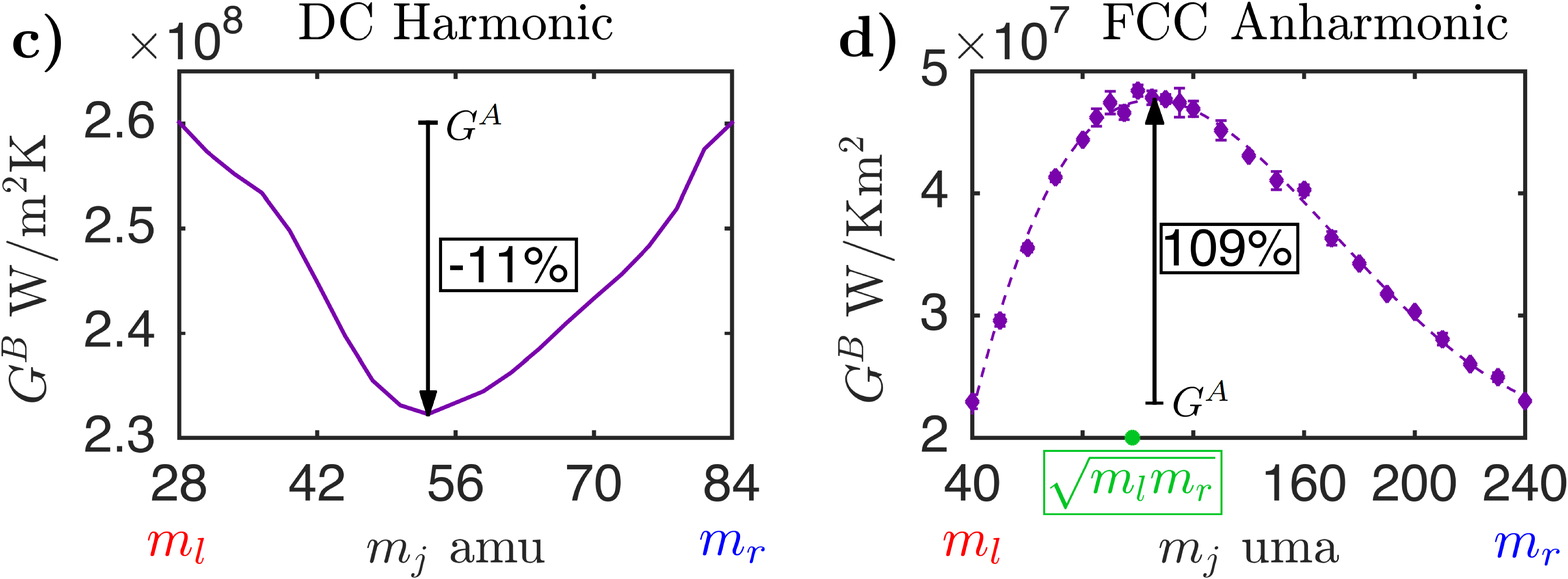}
	\caption{Conductance of bridged interfaces as the atomic mass of the intermediate layer or junction is varied between the contact masses. The conductance of the abrupt interface is indicated by the tail of the arrow. In the harmonic limit, a FCC crystal shows a relative enhancement of conductance from abrupt to bridged interface {\bf a)}, while a DC crystal shows the opposite {\bf c)}. In the anharmonic regime {\bf b)}, the conductance enhancement of a FCC crystal is three times larger than that of the harmonic limit {\bf a)}. Moreover, the maximum enhancement happens when the junction mass is close to the geometric mean of the contact masses ({\bf c)} $m_r=120$ amu and {\bf d)} $m_r=240$ amu). The dashed lines are fourth order polynomial functions that fit the NEMD data.}
	\label{fig_G_har_vs_anh}
\end{figure}

Figure~\ref{fig_G_har_vs_anh} plots the harmonic and anharmonic thermal conductances $G^B$ across interfaces with an added intermediate bridging layer or junction, belonging to a face centered cubic (FCC) crystal structure in one case and diamond cubic (DC) in the other. The boundaries between adjacent materials are assumed to be perfectly abrupt and the thickness of the junction is taken to be $6$ conventional unit cells. For each system, we vary the atomic mass of the junction $m_j$ in between the contact atomic masses $m_l$ and $m_r$. We assume that the crystal structure, lattice constant $a$ and interatomic force constants are invariant along the system, so we can isolate the effect of a change in atomic mass. Some consequences of relaxing these assumptions are discussed at the end of each of the following sections. The conductance in the harmonic regime is calculated from Landauer formalism in the classical limit using NEGF to obtain $MT$, while the conductance in the anharmonic regime is calculated from NEMD. Note that we report the conductance measured from the left to the right material including the contribution from the junction. Thus, the abrupt interface conductance from NEMD is larger than that of the bridged interface when the junction mass is equal to one of the contact masses. Those conductance values from Landauer formalism are equal because the calculations are harmonic. The details of the simulations are spelled out in Appendix \ref{app_sim_details}. 

Figures~\ref{fig_G_har_vs_anh}a and \ref{fig_G_har_vs_anh}b suggest that anharmonicity plays a key role in the relative enhancement of conductance from an abrupt (superscript $A$) to a bridged (superscript $B$) interface. The anharmonic simulations show a relative increase in conductance ($\Delta G=(G^B-G^A)/G^A=23\%$ at $\boldsymbol{T}=30$ K) three times larger than that of the harmonic simulations ($\Delta G=8\%$ at $\boldsymbol{T}=0$ K). This difference can not be explained in terms of the usual linear increase of conductance with temperature shown by NEMD simulations of abrupt interfaces (Fig.~\ref{fig_DeltaGmaxvsT_LJ}a) \cite{Stevens2007,Landry2009,Le2016}. In fact, the maximum conductance of bridged interfaces increases non linearly with temperature (Fig.~\ref{fig_DeltaGmaxvsT_LJ}a), with a rapid growth at low temperatures. This suggests the existence of a mechanism that limits the conductance enhancement just in the harmonic regime. In Section~\ref{secharmonic}, we explain that the limiting mechanism arises from the conservation of phonon energy and transverse momentum, which constrains the number of available transport channels across the interface. We also show that for certain crystal structures, this mechanism can even destroy the conductance enhancement of bridged interfaces over abrupt interfaces (Fig.~\ref{fig_G_har_vs_anh}c).

In the limit of zero temperature, the conductances calculated from Landauer and NEMD methods are in excellent agreement (Fig.~\ref{fig_DeltaGmaxvsT_LJ}a). The Landauer conductance is defined using the temperature drop between contacts at thermal equilibrium $G_L=q/\Delta \boldsymbol{T}_c$. Therefore, it {\it includes} additional resistances at the contacts that arise from the implicit scattering assumed to bring the distribution of phonons back to equilibrium (Appendix~\ref{App_CR}). On the other hand, the NEMD conductance is defined using the temperature drop right at the interface $G_{MD}=q/\Delta \boldsymbol{T}_i$ (Fig.~\ref{fig_DeltaGmaxvsT_LJ}b), so it {\it excludes} the resistances at the contacts. Those resistances cause the temperature drops at the boundaries of the heat baths (Fig.~\ref{fig_DeltaGmaxvsT_LJ}b), where thermal equilibrium is enforced. When we include the contact resistances into the NEMD conductance (right hand side of Eq.~\ref{equGLGMD}), we get the Landauer conductance
\begin{equation}
G_{L}=\lim_{\boldsymbol{T}\rightarrow 0}G_{MD}\frac{\Delta \boldsymbol{T}_i}{\Delta \boldsymbol{T}_c}.
\label{equGLGMD}
\end{equation}
Figure~\ref{fig_DeltaGmaxvsT_LJ}a shows an example of the excellent agreement of the two conductances once we account for the effects of the contact resistances. 

In the anharmonic regime, our simulations show that the conductance enhancement is maximum when the junction mass is close to the geometric mean of the contact masses $m_j\approx\sqrt{m_lm_r}$ (Figs.~\ref{fig_G_har_vs_anh}b and \ref{fig_G_har_vs_anh}d). This result is a consequence of the boundary resistance being an increasing function of the mass ratio of the materials at either side of the boundary (Sec.~\ref{secanharmonic}). Therefore, the sum of the boundary resistances is minimum when the ratio of the masses is equal ($m_j/m_l=m_r/m_j \rightarrow m_j\approx\sqrt{m_lm_r}$). Notably, this is a much more general result than an antireflection coating, which requires in addition a quarter wave plate to not just minimize but completely eliminate the sum of the boundary resistances through destructive interference, that only works at a single frequency for a homogeneous layer material.

\section{Harmonic limit: Increasing Transmission vs. Decreasing Conserving Modes} \label{secharmonic}

When phonons transit without interacting with each other, adding an intermediate layer does not necessarily increase the interfacial conductance (Fig.~\ref{fig_G_har_vs_anh}c). To understand this result, we start by rewriting Eq.~\ref{equGQ} to highlight the role of phonon transmission vs. number of transport channels on the interfacial conductance.  $G$ is related to the factor $MT$, which represents the sum over all the possible phonon transmissions between modes of the left contact, junction and right contact. Due to the perfectly abrupt nature of the boundaries, the system is periodic in the transverse direction, so that successful transmissions must conserve the transverse wavevector ($k_\perp$). We rewrite $MT$ to highlight the factors contributing to transport as
\begin{equation}
MT=\sum_{T_{k_\perp}\neq 0}T_{k_{\perp}}= M_c \left[\frac{1}{M_c}\sum\limits_{T_{k_\perp}\neq 0}T_{k_{\perp}}\right]=M_cT_c,
\label{equMcTc}
\end{equation}
where $k_\perp$ varies over the transverse Brillouin zone, the conserving modes $M_c$ counts the number of nonzero transmissions or transport channels across the interface, and $T_c$ is the average transmission over the conserving modes \cite{Polanco2015}. Using these definitions, we can rewrite the conductance as 
\begin{equation}
G= G_{M_c}\langle T_c\rangle_\omega,
\end{equation}
with the contribution to the conductance by the conserving modes $G_{M_c}$ given by
\begin{equation}
G_{M_c}=\frac{1}{A}\int_0^\infty\frac{d\omega}{2\pi}\hbar\omega\frac{\partial N}{\partial \boldsymbol{T}} M_c\xrightarrow{\hbar\rightarrow 0} \frac{k_B}{2\pi A}\int_0^\infty d\omega M_c
\end{equation}
and $\langle T_c\rangle_\omega=G/G_{M_c}$.

To calculate $M_c$ numerically, we find the propagating modes of the bulk left contact ($M_l$), junction ($M_j$) and right contact ($M_r$) by calculating $MT$ from NEGF for each homogeneous material, where $T_{k_\perp}=1$ for each mode and 0 otherwise. Then, the conserving modes are computed from 
\begin{equation}
M_c(\omega)=\sum_{k_\perp}\min\left[M_l(\omega,k_\perp),M_j(\omega,k_\perp),M_r(\omega,k_\perp)\right]
\label{equMc}
\end{equation}
Note that $M_c$ is a concept similar in spirit to the diffuse mismatch model \cite{Swartz1989}, since it depends only on the bulk properties of each individual material. Also note that we are assuming that tunneling across the junction is negligible, which is reasonable for junctions larger than four atomic layers. This assumption allows us to consider only transmissions involving propagating channels of the junction.

\begin{figure}[ht]
	\centering
	\includegraphics[width=86mm]{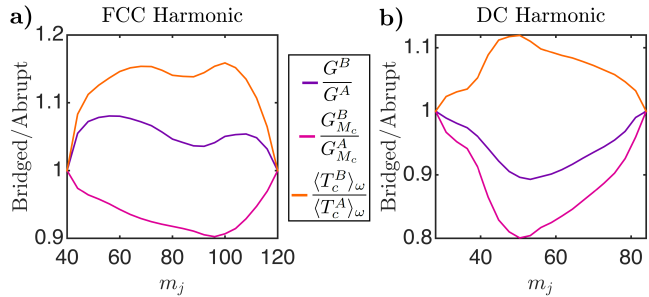}
	\caption{In the harmonic limit, the conductance of a bridged interface results from an interplay between increasing the transmission $\langle T_c\rangle_\omega$ due to decreasing the ``mismatch'' at each boundary and decreasing the number of conserving modes (due to a new restriction on the conservation of momentum coming from the intermediate material.}
	\label{fig_interfacewithadesionlayer}
\end{figure}

The relative enhancement in the conductance of a bridged (superscript $B$) interface compared to that of an abrupt (superscript $A$) interface depends on the interplay between increasing the transmission and decreasing the conserving modes. Figure~\ref{fig_interfacewithadesionlayer} compares the relative change in conductance, conserving modes and transmission using
\begin{equation}
\frac{G^B}{G^A}=\left[\frac{G^B_{M_c}}{G^A_{M_{c}}}\right] \left[\frac{\langle T^B_c\rangle_\omega}{\langle T^A_{c}\rangle_\omega}\right],
\end{equation}
with $M_{c}$ for the abrupt interface defined analogous to Eq.~\ref{equMc}, but the minimum is taken only over the contact modes. For the FCC crystal (Fig.~\ref{fig_interfacewithadesionlayer}a), the increase in transmission is enough to counter balance the decrease in modes. However, for the DC crystal (Fig.~\ref{fig_interfacewithadesionlayer}b), the decrease in  modes dominates and pushes the conductance of the bridged interface below that of the abrupt interface. The interplay between transmission and modes is a competition between increasing the value of individual transmitting channels vs. decreasing the number of them. 

\begin{figure}[ht]
	\centering
	\includegraphics[width=86mm]{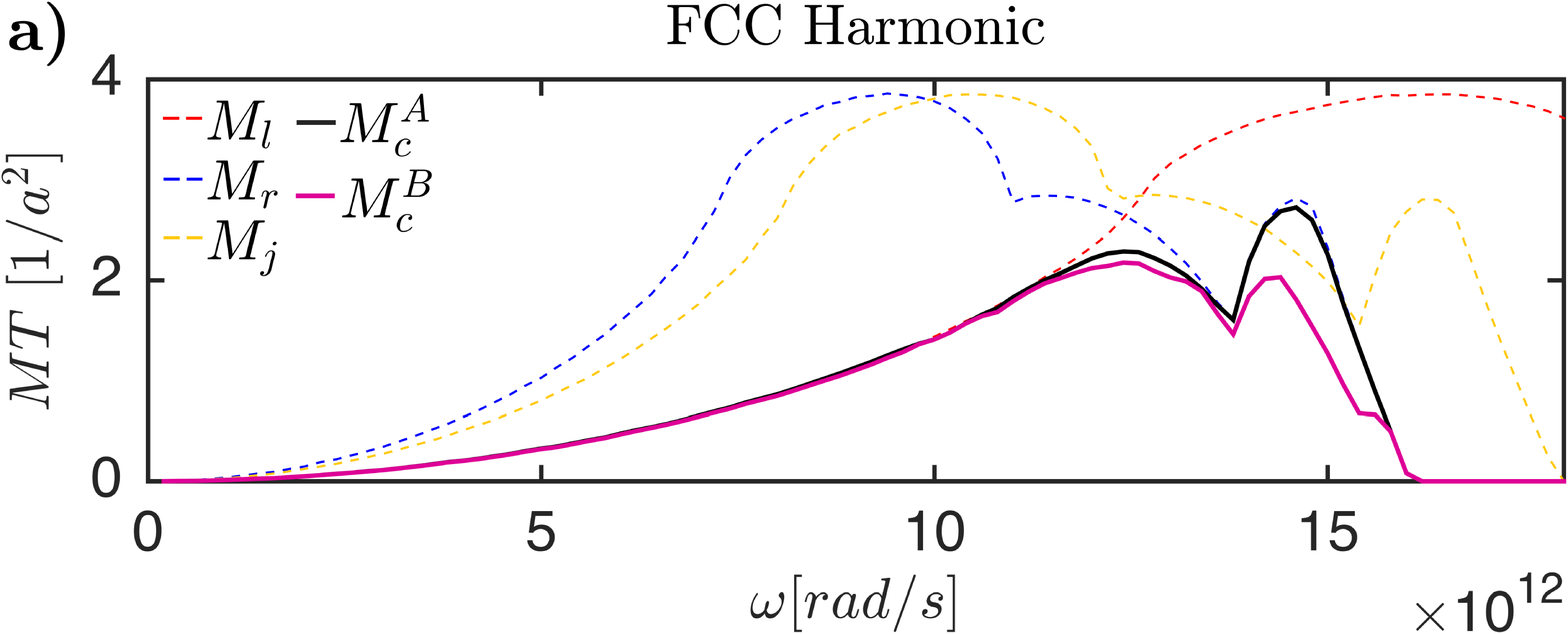}
	\includegraphics[width=86mm]{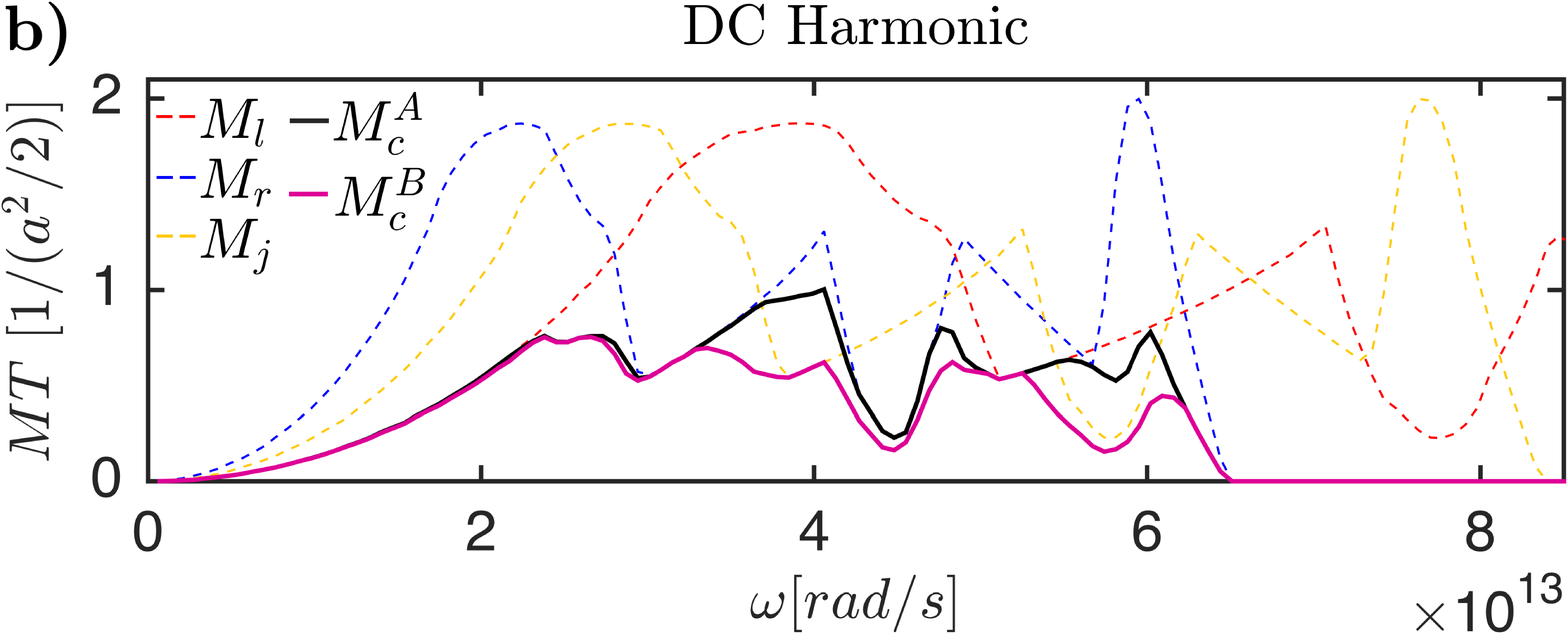}
	\caption{Available modes for the contacts and junction, and conserving modes of the bridged and abrupt interface. Adding the junction puts an extra constraint on the conserving modes that hurts $M_c$ and decreases the number of available modes. We plot the cases in which $G_{M_c}$ is minimum: $m_j=96$ amu for FCC crystal and $m_j=50.4$ amu for the DC crystal.}
	\label{fig_materialmodes}
\end{figure}

We can design an intermediate bridging layer at an abrupt interface to improve the impedance matching and increase the mode averaged transmission; however, it is important to note that an added layer always decreases the number of modes available for transport. This is a consequence of the need to conserve phonon energy and transverse momentum in three materials instead of two, which implies taking the minimum over three quantities instead of two (Eq.~\ref{equMc}). The extra constraint is more noticeable around frequencies where $M_j<\min(M_l,M_r)$. For instance, Fig.~\ref{fig_materialmodes}b shows a reduction of the conserving modes of the bridged interface relative to those of the abrupt interface around $4$ and $6\times10^{13}$ rad /s. Note that at low frequencies, the acoustic branches of the lightest material dominate the conserving modes and $M_c^A\approx M_c^B$. Thus, at low temperatures we expect $G^B>G^A$ for both crystal structures, FCC and DC, since $\langle T^B_c\rangle_\omega>\langle T^A_c\rangle_\omega$.

As long as the system remains periodic in the transverse direction, i.e. invariant lattice constant and perfectly abrupt boundaries between adjacent materials, the concepts developed in this manuscript apply. However, when the transverse symmetry assumption is relaxed, for example when the lattice constants are not the same or when there are random defects or interatomic mixing at the interface, phonons can change their momentum when they cross the interface. Therefore, the conserving modes do not represent anymore the available transport channels or modes across the interface. The number of transport channels is intimately related to the properties of the individual boundaries between adjacent materials, like the degree of interatomic mixing. In that case, conservation of energy allows us to define an upper bound for $MT$ \cite{Polanco2015}, 
\begin{equation}
MT^B(\omega)\leq\min(M_l(\omega),M_j(\omega),M_r(\omega))=M_{min}^B,
\end{equation} 
which can be used as a measure of the number of transport channels. Similar to the conserving modes, the minimum of the modes always decreases when a junction is added to an abrupt interface, because we are taking the minimum of three quantities instead of two (Fig.~\ref{fig_materialmodes}).

The conserving modes $M_c$ and minimum of the modes $M_{min}$ can be convenient starting points to look for junction materials that could enhance interfacial conductance. For instance, the combinations of materials that maximize $M_c$ or $M_{min}$ should be more amenable to increases in interfacial conductance.

\section{Anharmonic limit: Decreasing Boundary Resistance vs. Increasing Junction Resistance} \label{secanharmonic}

When phonons interact with each other during transport across the junction, for instance through anharmonic terms in the channel potential, they change their energy and momentum. This process relaxes the conservation constraints in the harmonic limit and decouples the system resistance as the sum of boundary resistances plus a junction resistance
\begin{equation}
R=R_{lj}+R_{j}+R_{jr}.
\label{equsumR}
\end{equation}
Figs.~\ref{fig_G_har_vs_anh}b and \ref{fig_G_har_vs_anh}d suggest that the maximum conductance, or minimum resistance, happens when the junction mass is close to the geometric mean of the contact masses. A similar result in terms of impedances was found for the analog one dimensional system, where phonon transport was elastic but incoherent \cite{Polanco2014}. The key element behind the result was that each boundary resistance is an increasing function of the impedance ratio of the materials at either side of the boundary. Thus, minimizing the sum of resistances requires equating the impedance ratios ($Z_j/Z_l=Z_r/Z_j \rightarrow Z_j=\sqrt{Z_lZ_r}$). Inspired by this result, it is tempting to suggest that $G$ is a function of the mass ratio alone. Unfortunately this is not true because heavier materials yield 
lower conductances due to their smaller cut off frequencies, which can be seen by rewriting Eq.~\ref{equGQ} as 
\begin{equation}
G=\frac{k_B}{2\pi A}\omega_{min}\langle MT\rangle_\omega,
\label{equGQ2}
\end{equation} 
with $\omega_{min}$ the minimum cut off frequency of the contacts $\omega_{min}=\min(\omega_{cl},\omega_{cr})$ and
\begin{equation}
\left\langle MT \right\rangle_\omega=\frac{1}{\omega_{min}}\int_0^{\infty}d\omega MT.
\label{equAvgMTw}
\end{equation}

\begin{figure}[ht]
	\centering
	\includegraphics[width=86mm]{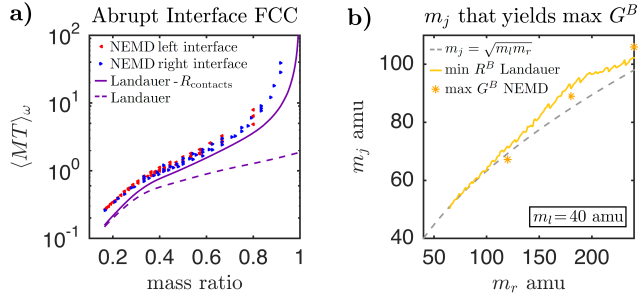}
	\caption{{\bf a) }$\langle MT\rangle_\omega$, plotted per conventional unit cell, is a function of the mass ratio for abrupt interfaces. {\bf b)} Junction mass $m_j$ that leads to minimum resistance vs. right contact mass $m_r$ while keeping the left contact mass fixed. The mass follows closely the geometric mean of the contact masses.}
	\label{fig_AvgMT}
\end{figure}

Although the boundary conductance is not a function of the mass ratio alone, the conductance over the frequency cut-off $G/\omega_{min}$ is. Figure~\ref{fig_AvgMT}a shows that $\langle MT\rangle_\omega$ is only a function of the mass ratio for both the anharmonic and harmonic limits. The anharmonic data results from combining Eq.~\ref{equGQ2} with the boundary conductance extracted from our NEMD simulations. The data from the boundary between left contact and junction (red triangles) is a little larger than that from the boundary between junction and right boundary (blue triangles) because the temperature is greater at the left boundary. The harmonic data comes from the Landauer conductance after we have subtracted the contact resistances (Appendix~\ref{App_CR}). If that is not the case, we obtain the dashed line in Fig.~\ref{fig_AvgMT}a, which is bounded for unity mass ratio due to the resistances at the contacts. 

Replacing the boundary resistances ($R=1/G$) from Eq.~\ref{equGQ2} into Eq.~\ref{equsumR} and using $\langle MT\rangle_\omega$ from Landauer (solid line in Fig.~\ref{fig_AvgMT}a), we numerically find the junction mass that maximizes interfacial conductance as a function of the ratio of the contact masses (solid line Fig.~\ref{fig_AvgMT}b). The noise in the plot is caused by the interpolation error in $\langle MT\rangle_\omega$. The fair agreement of this curve with the results from NEMD simulations (Figs.~\ref{fig_G_har_vs_anh}b and d) suggests that the knowledge of the harmonic boundary conductance is enough to approximate the junction mass that maximizes the conductance. Nevertheless, one of the reasons behind the discrepancy is the flattening of the curve around the peak (Figs.~\ref{fig_G_har_vs_anh}b and d, Appendix~\ref{app_sim_details} and Fig.~\ref{fig_maxG_MDvsNEGF}), which combined with the uncertainty of the NEMD results produces a corresponding spread in the maximum. In fact, the large spread of the conductance maximum and its relative insensitivity around that point with changes of junction mass is quite convenient for engineering, as it widens our choice of bridging masses that yield an overall large conductance. 

The solid curve on Fig.~\ref{fig_AvgMT}b follows closely the geometric mean of the contact masses (dashed line) due to the dominant dependance of $G$ on the mass ratio. The deviations from this mean arise from the dependence of G on the overall phonon cut-off frequency $\omega_{min}$, which adds a more complicated mass dependence. We can better understand the trend of maximum conductance by minimizing Eq.~\ref{equsumR}
\begin{equation}
\frac{\partial R}{\partial m_j}= \frac{2\pi A_{cu}}{k_B\omega_{cr}}\left[ \frac{F_{lj}}{2\sqrt{m_jm_r}} \right. - \sqrt{\frac{m_j}{m_r}}\underbrace{\frac{m_l}{m_j^2} F'_{lj}}_{\text{term } \alpha} + \underbrace{ \left. \frac{1}{m_r}F'_{jr}  \right]}_{\text{term } \beta} =0.
\label{equminR}
\end{equation}
To obtain Eq.~\ref{equminR} we use $m_l<m_j<m_r$, Eq.~\ref{equGQ2}, $F_{lj}=\langle MT(m_l/m_j) \rangle^{-1}$, $F_{jr}=\langle MT(m_j/m_r) \rangle^{-1}$, $F'$ the derivative of $F$ with respect to $m_j$ and we neglect $\partial R_j/\partial m_j$. We also express the cut off frequencies of the junction $\omega_{cj}$ and right contact $\omega_{cr}$ in terms of the cut off frequency of the left contact ($\omega_{cj}=\omega_{cl}\sqrt{m_l/m_j}$, $\omega_{cr}=\omega_{cl}\sqrt{m_l/m_r}$). This is possible because the materials are identical except for the atomic mass, so the dispersion is a copy of the same function expanded or contracted along the frequency axis. $A_{cu}$ is the area of the conventional unit cell that converts the value of $\langle MT\rangle_\omega$ per conventional unit cell in Fig.~\ref{fig_AvgMT}a to per meter squared. When the ratio between the contact masses is close to one, choosing the junction mass close to the geometric mean of the contact masses maximizes the conductance (Fig.~\ref{fig_AvgMT}b). In that case, $\sqrt{m_j/m_r}\approx 1$, $F_{lj}\approx 0$ and Eq.~\ref{equminR} reduces to the terms $\alpha$ and $\beta$. This expression is minimum when $m_j\approx \sqrt{m_lm_r}$. As the ratio of the contact masses increases, the junction mass that maximizes the conductance remains close to the geometric mean. This happens because the deviation of the second term relative to the term $\alpha$, caused by $\sqrt{m_j/m_r}< 1$,  is balanced to some extent by the increase of $F_{lj}$ on the first term (note that $F'$ is negative). 

To minimize the resistance, 1) we assume the boundary resistances are in series, 2) we show the boundary conductance is an increasing function of the mass ratio, and 3) we conclude that the minimum resistance happens when the mass ratios are equal. This strategy can be used beyond systems with perfect boundaries where only the mass is allowed to change. We expect the same minimization outcome, $m_j\approx\sqrt{m_lm_r}$, when interatomic mixing is added at the boundaries. Mixing can ether suppress \cite{Hopkins2013} or enhance \cite{English2012, Tian2012, Polanco2015} each boundary conductance. Either way, we still expect a similar increasing trend of $\langle MT\rangle_\omega$ with mass ratio dictated by the frequency minimum of the modes $M_{min}$ instead of the conserving modes $M_c$ \cite{Polanco2015}. By analogy, if we only allow changes of the interatomic force constants by varying the $\epsilon$ parameter of the Lennard-Jones potential, we expect minimum resistance when $\epsilon_j\approx\sqrt{\epsilon_l\epsilon_r}$. Although the force constants and masses have opposite effects on the cut off frequency, we still expect $\langle MT\rangle_\omega$ to be an increasing function of the $\epsilon$ ratio. When we allow changes of both the masses and force constants, a similar analysis suggests that the minimum resistance happens when $m_j/\epsilon_j\approx\sqrt{m_lm_r/\epsilon_l\epsilon_r}$. Further studies are necessary to confirm our hypotheses and to extend it to matrix versions of $m$ and $\epsilon$ for anisotropic systems. 

We expect further enhancement of the conductance between the contacts by stacking several intermediate thin layers whose atomic masses change in an exponential fashion.  This result follows from choosing the mass of each layer as the geometric mean of the masses of the adjacent layers. Each geometric mean choice minimizes the sum of the boundary resistances adjacent to a particular layer. A similar conjecture has been demonstrated for 1D incoherent systems \cite{Polanco2015}.

\section{Conclusion}

We study the enhancement of thermal conductance when a thin film layer or junction is inserted at an abrupt interface. Our simulations show three times larger enhancement when the harmonic approximation is relaxed, which highlights the important role of phonon-phonon interactions in this transport process. In fact, in the harmonic limit, adding a junction to the abrupt interface does not necessarily enhance the conductance. The result depends on the interplay between 1) increasing the transmission by improving the ``matching'' of the contacts and 2) decreasing the number of available transport channels that conserve energy and transverse momentum. When anharmonicity kicks in, the conservation constraints are relaxed and the resistance of the system can be split into the sum of boundary resistances. The resistance is minimized when the junction mass is the geometric mean of the contact masses, which follows from the increasing trend of the boundary resistance with mass ratio. The strategy to find the maximum conductance can be used beyond systems with perfect interfaces where only mass is changing. We hypothesize that for a graded junction the geometric mean result generalizes to an exponential progression of masses that can push the enhancement beyond that of a single layer. This paper exemplifies the powerful combination of Landauer and NEMD to study the harmonic vs. the anharmonic contributions to thermal conductance.

\begin{acknowledgments}
C.A.P., J.Z. and A.W.G. are grateful for the support from NSF-CAREER (QMHP 1028883) and from NSF-IDR (CBET 1134311). This work used the Extreme Science and Engineering Discovery Environment (XSEDE) \cite{XSEDE} (DMR130123), which is supported by National Science Foundation grant number ACI-1053575. R.R and P.M.N. acknowledge the financial support of the Air Force Office of Scientific Research (Grant No. FA9550-14-1-0395).
\end{acknowledgments}

\appendix

\section{Simulation Details} \label{app_sim_details}

\begin{figure}[ht]
	\centering
	\includegraphics[width=86mm]{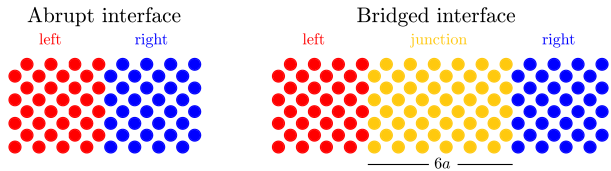}
	\caption{Lateral view of the abrupt and bridged interfaces simulated in this work. Each ball represents a primitive unit cell.}
	\label{fig_Systems}
\end{figure}

We calculate the thermal conductance between the left and right contacts of abrupt and bridged interfaces (Fig.~\ref{fig_Systems}). For each system, the crystal structure, lattice constant $a$ and interatomic force constants are invariant. The boundaries between adjacent materials are perfectly abrupt. The junction is six conventional unit cells long and the junction atomic mass $m_j$ is varied between the contact atomic masses $m_l$ and $m_r$. We simulate interfaces on FCC and DC crystal structures. 
For the DC interfaces, the interatomic force constants are calculated from the Stillinger-Weber interatomic potential for Si \cite{Stillinger1985}. This potential describes the energy in terms of two and three body potentials and includes interactions up to the second-nearest neighbors. The equilibrium lattice constant for this structure is $a=5.431$ \AA\ at $\boldsymbol{T}=0$ K. The mass for the left contact is chosen as the silicon mass $m_l=28$ amu and the right contact mass is chosen as $m_r=84$ amu, which is close to the mass of germanium. The conductance  for the abrupt interface calculated from Landauer or harmonic NEGF in the classical limit is $G^A=260.1$ MW m$^{-2}$ K$^{-1}$ (Fig.~\ref{fig_G_har_vs_anh}c). This value is in good agreement with reported conductance values $G^A = 276.6$ MW m$^{-2}$ K$^{-1}$ at $\boldsymbol{T}=300$ K \cite{Polanco2015} and $G^A = 280$ m$^{-2}$ K$^{-1}$ \cite{Tian2012}, which belong to abrupt interfaces with contact masses $m_l=28$ amu and $m_r=72$ amu. Our value is smaller because of the heavier mass on the right contact, which reduces the available phonon spectrum for conduction. 

For the FCC interfaces, the interatomic force constants are calculated from the Lennard-Jones potential with parameters $\epsilon=0.0503$ eV, $\sigma=3.37$ \AA, and a cutoff distance of $2.5\sigma$. This potential includes interactions up to the fifth-nearest neighbors and is chosen to be identical to that used by English {\it et al.} \cite{English2012} to have a point of reference for benchmarking. The equilibrium lattice constant for this structure is $a=5.22$ \AA\ at $\boldsymbol{T}=0$ K. The mass of the left contact is fixed to $m_l=40$ amu, while the mass of the right contact is varied from 40 amu to 240 amu. For Fig.~\ref{fig_G_har_vs_anh}a and b, Fig.~\ref{fig_DeltaGmaxvsT_LJ}, Fig~\ref{fig_interfacewithadesionlayer}a and Fig.~\ref{fig_materialmodes}a, $m_r=120$ amu. For Fig.~\ref{fig_G_har_vs_anh}d, $m_r=240$ amu.  From harmonic NEGF in the classical limit, the conductance for the abrupt interface is $G^A=59.3$ MW m$^{-2}$ K$^{-1}$. This is in excellent agreement with the conductance from NEMD at $\boldsymbol{T}=2$ K including the contact resistances $G^A=61.0$ MW m$^{-2}$ K$^{-1}$.

For the NEGF simulations, we take advantage of the transverse symmetry of the systems. We calculate $MT$ in transverse wave-vector space ($k_\perp-$ space) to simplify the 3D problem into a sum of 1D independent problems. The transverse Brillouin zone was split into $50\times 50$ grid points for both the FCC and DC crystals.

For the NEMD simulations we use the LAMMPS MD simulator on a system with $10\times10\times62$ conventional unit cells and a time-step of 2 fs. We impose periodic boundary conditions over $x$ and $y$ directions and set the atomic layers at the two ends of the system as walls. Heat is added to the system from the left edge and removed from the right edge using the Langevin thermostat. The baths temperatures are set to $\boldsymbol{T}_{\text{bath}}=(1\pm0.1)\boldsymbol{T}$ with a time constant of 1.07 ps over blocks of 10 unit cells length. This setup for the thermostat is done to ensure sufficient phonon-phonon scattering that prevents size effects. On the computations at very low temperatures ($\boldsymbol{T}=2$ K), we test for size effects by changing the cross section to $15\times15$ and $20\times20$. We also vary the length of the domain to 100 unit cells and decrease the thermostat time constant to 0.54 ps. No significant change in the thermal boundary conductance is noticed. 

To prevent changes of pressure as the temperature varies from affecting thermal transport at the interface, we account for the thermal expansion of the system. We perform equilibration runs under zero pressure at different temperatures using the isothermal-isobaric ensemble (NPT). The results are used to find the dependence of the lattice constant with temperature, which is fitted to a third order polynomial function:
\begin{equation}
a(\boldsymbol{T})=5.2222+0.0004\boldsymbol{T}+10^{-6}\boldsymbol{T}^2-4\times 10^{-9}\boldsymbol{T}^3 \ \text{\AA}.
\end{equation}

Atoms are first equilibrated under the microcanonical ensemble (NVE) for 4 ns. Next, heat is added to the system for 10 ns to achieve steady state. Then, the temperature is recorded for 6 ns to ensure a proper statistical average. From the temperature profile, we estimate the thermal boundary conductance dividing the heat flux over the temperature drop, which arises from a linear fit of the temperature at each lead extrapolated to the interface.

Most of the conductance values from NEMD reported in this paper are averages over five independent calculations whose initial condition is generated randomly. The maximum conductances reported as asterisks on Fig.~\ref{fig_AvgMT}b are the maximum of fourth order polynomial functions used to fit the NEMD data. 

\begin{figure}[ht]
	\centering
	\includegraphics[width=86mm]{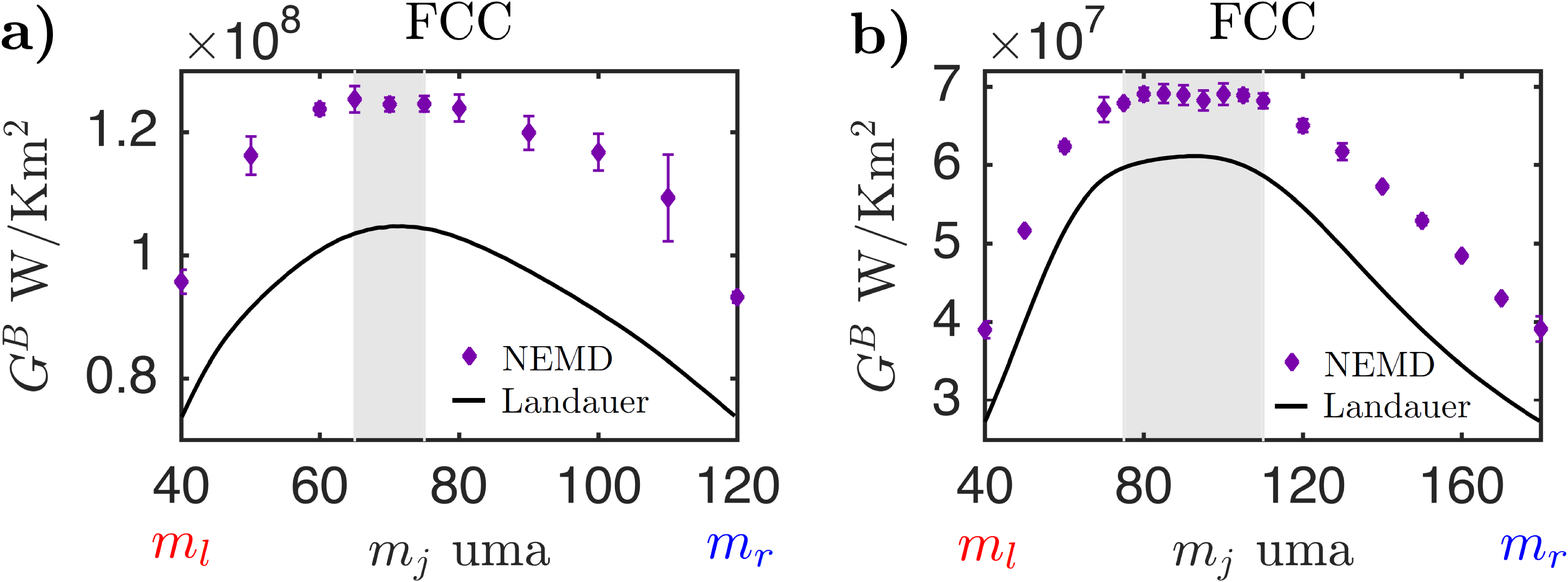}
	\includegraphics[width=86mm]{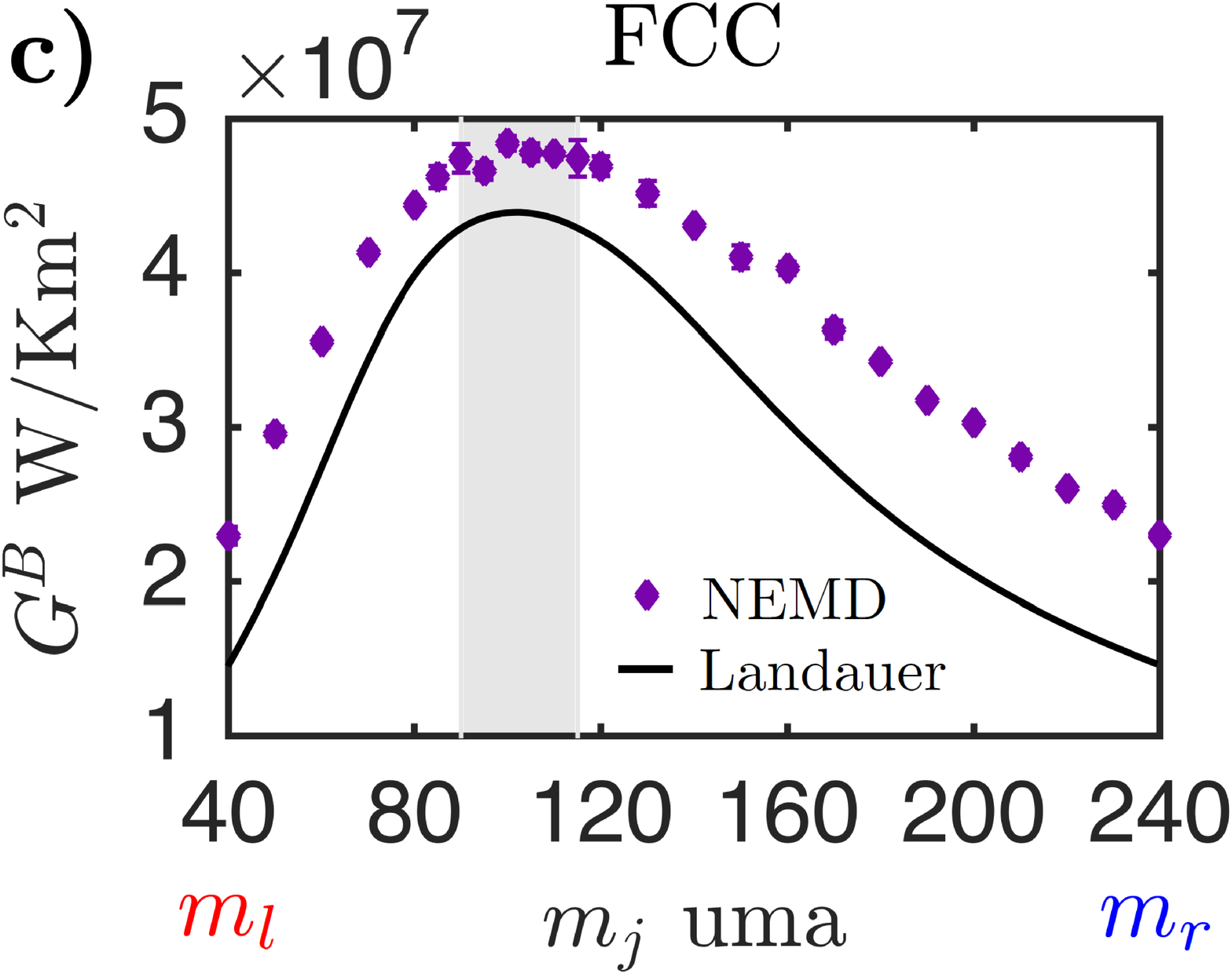}
	\caption{Conductance vs. junction mass for different right contact masses. {\bf a)} $m_r=120$ amu, {\bf b)} $m_r=180$ amu, {\bf c)} $m_r=240$ amu. The shaded area shows the masses whose NEMD conductance enhancement is within 5\% of the maximum enhancement. The Landauer curves come from Eq.~\ref{equsumR}$-R_j$ using $\langle MT\rangle_\omega$ from Landauer (solid line Fig.~\ref{fig_AvgMT}b).}
	\label{fig_maxG_MDvsNEGF}
\end{figure}

The discrepancy in Fig.~\ref{fig_AvgMT}b between the maximum conductance of the bridged interface extracted from NEMD and the one predicted from Landauer can be attributed to the flattening of the $G$ vs. $m_j$ curve around the maximum. Figure~\ref{fig_maxG_MDvsNEGF} shows the region (shaded area) where the enhancement of conductance is within 5\% of the maximum enhancement. In that region, it is difficult to pinpoint the exact location of the maximum due to the uncertainty of the NEMD results. Nevertheless, the overall shapes of the Landauer and NEMD curves used to predict the maximum are in excellent agreement (Fig.~\ref{fig_maxG_MDvsNEGF}). The different height between the curves is a consequence of the larger boundary conductances obtained with NEMD simulations (Fig.~\ref{fig_G_har_vs_anh}b). The Landauer curve does not include the intrinsic resistance of the junction. Thus, its similarity with the shape of the NEMD curve in Fig.~\ref{fig_maxG_MDvsNEGF} suggests that individual boundaries play a dominant role in the maximization process. Thus, the quality of the conductance enhancement depends mostly on our ability to decrease the sum of the boundary resistances.

\section{Contact Resistance}
\label{App_CR}

According to Landauer theory, the conductance of a device in between two contacts at thermal equilibrium is defined by Eq.~\ref{equGQ}. When the device and the contacts are made of the same material, the transmission $T$ equals one. In that case, we get the upper limit of conductance, which is proportional to the quantum of conductance times the number of propagating channels \cite{Datta2005}. Since $T=1$, there can not be any resistance associated with the flow of phonons inside the device. Therefore, the maximum conductance measures the resistance at the contacts. This resistance arises from the implicit scattering processes that have to happen at the contacts, to bring the flowing phonons back to an equilibrium distribution \cite{Datta2005}. The diffusive nature of those scattering processes allows us to split the resistance associated with the maximum conductance into the sum of the resistances at the contacts. Since for a homogeneous material those resistances should be equal, we can define the contact conductance in the classical limit as
\begin{equation}
G_c=\frac{2k_B}{2\pi A}\omega_{c}\langle M\rangle_\omega,
\label{equGQc}
\end{equation} 
with $\omega_{c}$ the cut off frequency of the material and
\begin{equation}
\left\langle M \right\rangle_\omega=\frac{1}{\omega_{c}}\int_0^{\infty}d\omega M.
\label{equAvgMTc}
\end{equation}

The conductance from Landauer theory $G_{L}$ is the parallel of the device conductance $G_{d}$ with the contact conductances $G_l$ and $G_r$, so the device conductance is given by
\begin{equation}
G_{d}=\frac{\left(\frac{G_lG_r}{G_l+G_r}\right)G_{L}}{\left(\frac{G_lG_r}{G_l+G_r}\right)-G_{L}}.
\end{equation}  
Combining $G_d$ with Eq.~\ref{equGQ2} we extract the $\langle MT\rangle_\omega$ referred as Landauer-$R_{\text{contacts}}$ in Fig.~\ref{fig_AvgMT}a. The method presented here is one way to approximate the contact resistances. Other approximations have been presented, which include an analogy to the four probe measurements \cite{Katerberg1977,Tian2012}.

From the temperature profile of the NEMD simulations when $\boldsymbol{T}\rightarrow 0$ (Fig.~\ref{fig_DeltaGmaxvsT_LJ}b), we can estimate the contact resistances, which are related to the temperature drops at the edges of the heat baths. At the contacts or heat baths, every time step the velocities of the atoms are rescaled to a thermalized distribution, which emulates phonon-phonon scattering processes bringing the region back to equilibrium. Everywhere else, phonons do not interact because the low temperature makes the system harmonic. Therefore, once a phonon leave the contacts it can not relax its energy creating a non equilibrium distribution everywhere outside the bath regions. The temperature plotted in Fig.~\ref{fig_DeltaGmaxvsT_LJ}b is a representation of the total kinetic energy of the region with an equilibrium distribution. 

The conductance measured from NEMD uses the temperature drop at the interface ($\Delta\boldsymbol{T}_{\text{i}}$), while the one measured from Landauer uses the temperature drop at the contacts ($\Delta\boldsymbol{T}_{\text{c}}$). Since the heat flux ($q=G\Delta\boldsymbol{T}$) crossing the system is the same, we can relate the conductances from the two methods with Eq.~\ref{equGLGMD}. Using this relation, we found excellent agreement between the results from Landauer and NEMD (Fig~\ref{fig_DeltaGmaxvsT_LJ}a). Another example supporting this relationship is shown in our recent work \cite{Le2016}.

%\bibliography{bib_har_vs_anh}% Produces the bibliography via BibTeX.


\end{document}